\documentclass{iopart}
\usepackage{graphicx}
\usepackage{iopams}
\usepackage{bm}
\newcommand{\eq}{{\,=\,}}

\begin{document}
\vspace*{-1cm}
\title[Extracting QGP viscosity from RHIC data]%
{Extracting the QGP viscosity from RHIC data \\
\ \ -- a status report from viscous hydrodynamics
}
\author{Huichao Song and Ulrich Heinz}
\address{Department of Physics, The Ohio State University, Columbus,
         OH  43210, USA}

\begin{abstract}
We report recent progress on causal viscous hydrodynamics for relativistic 
heavy ion collisions. For fixed specific shear viscosity $\eta/s$,  
uncertainties in the elliptic flow arising from initial conditions, 
equation of state, bulk viscosity and numerical viscosity, and the 
treatment of the highly viscous hadronic stage and freeze-out procedure 
are analysed. A comparison of current viscous hydrodynamic results with 
experimental data yields a robust upper limit 
$\eta/s < 5 \times \frac{1}{4\pi}$.
\\[-1cm]
\end{abstract}

\section{Introduction}
\label{sec1}
\vspace*{-2mm}

The viscosity of the quark-gluon plasma is presently a hotly debated
subject. Its computation from first principles is difficult. It is 
thus desirable to try extracting it from experimental data. Viscous
hydrodynamics provides a tool that can be used to attack this problem 
while simultaneously extending the region of applicability of the 
hydrodynamic approach beyond that of ideal fluid dynamics.

During the last years, several groups have independently developed
(2+1)-dimensional viscous hydrodynamic codes to describe the mid-rapidity
spectra from heavy-ion collisions at Relativistic Heavy Ion Collider 
(RHIC) energies. First results were published in Refs. 
\cite{Romatschke:2007mq,Song&HeinzPLB08,Dusling&Teaney:2007gi,%
Song&HeinzPRC08-1,Chaudhuri:2008sj,Song&HeinzPRC08-2,Romatschke:2008,%
Molnar:2008xj,Heinz:2008qm}. It was found that shear viscosity decelerates 
the longitudinal expansion, thereby initially slowing down the cooling 
process, but accelerates the transverse expansion, resulting in 
more radial flow and flatter hadron $p_T$-spectra. In non-central
collisions, the elliptic flow coefficient $v_2$ was found to be very 
sensitive to shear viscosity: given the large expansion rates of
heavy-ion collision fireballs, even minimal viscosity saturating
the KSS bound $\eta/s\geq1/4\pi$ \cite{Kovtun:2004de} for the viscosity 
to entropy density ratio can lead to a strong (and thus easily measurable) 
suppression of $v_2$. Assuming the availability of a well-established 
ideal fluid dynamical baseline for $v_2$ as a function of collision 
energy, centrality and system size, measurements of this suppression  
could thus be used to constrain the QGP shear viscosity from experimental 
data. Other observables (e.g. slopes of $p_T$-spectra) also depend
on $\eta/s$, but $v_2$ appears to show the strongest sensitivity,
so it has been studied most extensively, and we will focus on it. 
We will concentrate on viscous hydrodynamics, leaving a discussion
of more microscopic approaches (e.g. \cite{Xu:2004mz,Huovinen:2008te}) for 
another time.

The first results for viscous $v_2$ suppression published by the
different groups seemed to show large discrepancies, ranging from 
20\% to 70\% for 'minimal viscosity' $\eta/s=1/4\pi$. A detailed code 
verification process \cite{code-checking}, carried out within the TECHQM 
collaboration \cite{TECHQM}, eliminated the possibility that these 
differences were caused by numerical error. Systematic studies during 
the last few months revealed their physical origins and are summarized 
in this talk. Specifically, we will discuss the effects of system size, 
equation of state (EOS), and different versions of the Israel-Stewart 
equations used to evolve the viscous terms in the energy-momentum tensor, 
on viscous $v_2$ suppression. A combination of these effects has been 
shown to resolve the apparent discrepancies reported by different
groups \cite{Song&HeinzPRC08-2}. Further, we will discuss effects from 
different freeze-out procedures \cite{Dusling}, different initializations
(Glauber model vs. Color Glass Condensate approach) \cite{Romatschke:2008}, 
bulk viscosity \cite{Song&Heinz-bulkVis}, and numerical viscosity 
\cite{Song&HeinzPRC08-2}. Quantifying these effects is important for
understanding the uncertainties in extracting the shear viscosity to
entropy ratio from experimental data.

\section{Viscous hydrodynamics in 2+1 dimensions}
\label{sec2}
\vspace*{-2mm}

In this section, we briefly review the causal viscous equations
solved by the {\tt VISH2+1} code (Viscous Isreal-Stewart Hydrodynamics 
in 2+1 space-time dimensions) developed at OSU, used to simulate
QGP and hadronic matter expansion with exact longitudinal boost
invariance but arbitrary dynamics in the transverse directions 
\cite{Heinz&Song05}. For simplicity, we assume zero net baryon number 
and heat conductivity. {\tt VISH2+1} solves the equations for energy momentum 
conservation $d_m T^{mn}=0$, with
\begin{eqnarray}
 T^{mn} = e u^m u^n - (p+\Pi)\Delta^{mn} + \pi^{mn},\ \ \
 \Delta^{mn} =g^{mn}{-}u^m u^n,
\end{eqnarray}
together with kinetic equations for the viscous shear pressure
tensor $\pi^{mn}$ and the bulk pressure $\Pi$:
\begin{eqnarray}
\label{eq2}
\Delta^{mr}\Delta^{ns} D\pi_{rs}
=-\frac{1}{\tau_{\pi}}(\pi^{mn}{-}2\eta\sigma^{mn})
 -\frac{1}{2}\pi^{mn} \frac{\eta T}{\tau_\pi}
       d_k\left(\frac{\tau_\pi}{\eta T}u^k\right), 
\\ \label{eq3}
        D \Pi
=-\frac{1}{\tau_{\Pi}}(\Pi+\zeta \theta)
 -\frac{1}{2}\Pi\frac{\zeta T}{\tau_\Pi}
       d_k\left(\frac{\tau_\Pi}{\zeta T}u^k\right).
\end{eqnarray}
Here $D\eq u^m d_m$ is the time derivative in the local comoving
frame, $\nabla^m\eq\Delta^{ml}d_{l}$ is the spatial gradient in that
frame, and $\sigma^{mn}\eq\nabla^{\left\langle m\right.} u^{\left.n
\right\rangle}\eq\frac{1}{2}(\nabla^m u^n{+}\nabla^n
u^m)-\frac{1}{3} \Delta^{mn}\theta$ (with the scalar expansion rate
$\theta\equiv d_k u^k =\nabla_k u^k$) is the symmetric and traceless
velocity shear tensor. $d_m$ denotes the components of the covariant 
derivative in the curvilinear coordinates $(\tau,x,y,\eta_s)$ 
\cite{Heinz&Song05}. 

For systems with conformal symmetry the last terms in 
Eqs.~(\ref{eq2},\ref{eq3}) can be written in different forms
as discussed in \cite{Song&HeinzPRC08-2}. While these forms are 
equivalent for conformal systems, they differ in principle for systems 
with an EOS that breaks conformal invariance, e.g. through a phase 
transition. These differences turn out to be negligible in practice
\cite{Song&HeinzPRC08-2}. Without the last terms, Eqs.~(\ref{eq2},\ref{eq3})
are known as {\sl simplified I-S equations} whereas their complete
versions will be called {\sl full I-S equations}. In spite of this name,
these equations still do not include all possible second-order terms
in a gradient expansion around the ideal fluid (locally thermalized)
limit. For a conformal theory with vanishing chemical potentials 4 other
terms can be added to the right hand side of Eq.~(\ref{eq2}), with 
additional coefficients $\lambda_1$, $\lambda_2$, $\lambda_3$, and 
$\kappa$ ($\kappa=0$ in Minkowski space) \cite{Baier:2007ix}. These terms
include couplings to the vorticity tensor \cite{Baier:2007ix} which turns
out to be small in heavy-ion collisions if the initial longitudinal velocity
profile is boost invariant \cite{Romatschke:2007mq}. Even more terms arise 
in a kinetic theory derivation that does not assume conformal symmetry 
\cite{Betz:2008me}, including terms that couple Eqs.~(\ref{eq2}) and 
(\ref{eq3}). Their coefficients can be obtained from kinetic theory 
\cite{Betz:2008me,York:2008rr} at weak coupling or from the AdS/CFT
correspondence at strong coupling \cite{Baier:2007ix}. Still another
approach was developed by \"Ottinger and Grmela (see references in 
\cite{Dusling&Teaney:2007gi}); when translated into Israel-Stewart form,
it also makes specific predictions for the coefficients of these 
second-order terms \cite{Dusling}. The so far accumulated numerical 
evidence \cite{Song&HeinzPRC08-2,Romatschke:2008,code-checking} suggests 
that, except for the last terms in Eqs.~(\ref{eq2}) and (\ref{eq3}), all 
other second order terms are unimportant in practice, but a systematic
study that confirms this beyond doubt remains outstanding. 

The shear viscosity $\eta$, bulk viscosity $\zeta$ and the
corresponding relaxation times $\tau_\pi$ and $\tau_\Pi$ are free
parameters in {\tt VISH2+1}. For most of the published numerical simulations 
of (2+1)-d viscous hydrodynamics, the shear viscosity has been set to 
the KSS minimal value $\eta/s=1/4\pi\simeq 0.08$, and the bulk viscosity 
was set to zero, $\zeta/s=0$. Bulk viscous effects on the evolution
of elliptic flow in non-central collisions were recently studied by 
us \cite{Song&Heinz-bulkVis} and are here reported for the first time
(see Sec.~\ref{sec5}). For the relaxation times $\tau_\pi$ and $\tau_\Pi$
most authors used a constant multiple of $\eta/sT$. Physical
observables turn out to be largely insensitive to the value of $\tau_\pi$
if (and only if!) the last terms are included in Eqs.~(\ref{eq2}) and
(\ref{eq3}), i.e. the {\sl full} (and not the {\sl simplified!}) 
Israel-Stewart equations are used \cite{Song&HeinzPRC08-2} (see 
Sec.~\ref{sec3} below).

The equation of state (EOS), initial and final conditions are additional 
inputs for both ideal and viscous hydrodynamics. Details can be found in 
the original literature \cite{Dusling&Teaney:2007gi,Song&HeinzPRC08-1,%
Song&HeinzPRC08-2,Romatschke:2008} and will not be explained here. Two 
different EOS will be used in these proceedings: {\tt SM-EOS~Q} is a slightly 
smoothed version \cite{Song&HeinzPRC08-1} of {\tt EOS~Q} \cite{reviews} 
which describes a non-interacting massless QGP phase matched to a chemically
equilibrated massive hadron resonance gas (HRG) through a Maxwell 
construction. {\tt EOS~L} matches the HRG EOS below $T_c$ smoothly
with the lattice QCD EOS above $T_c$ \cite{Song&HeinzPRC08-1}. Our 
{\tt EOS~L} is close to but not identical with the quasiparticle EOS 
used by Romatschke \cite{Romatschke:2007mq,Romatschke:2008}.

\section{Viscous $\bm{v_2}$ suppression: effects from system size, EOS and 
different versions of Israel-Stewart equations}
\label{sec3}
\vspace*{-2mm}

In this section we will briefly discuss the different manifestations of 
shear viscosity when one varies system size and EOS and uses different
versions of the I-S equations \cite{Song&HeinzPRC08-2}. As mentioned in 
the Introduction, this analysis resolves the initially puzzling differences
between the results published by different groups.  

\begin{figure}[htb]
  \begin{center}
 \includegraphics[width=0.98\linewidth,clip=]{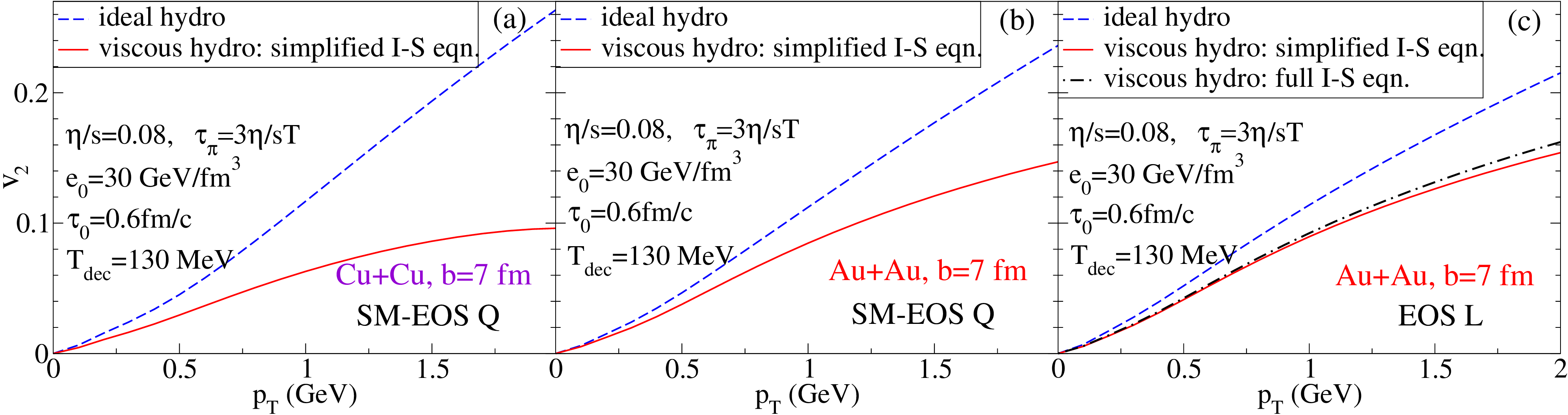}
  \end{center}
\vspace*{-2mm}
\caption{\label{F1}(Color online) Differential elliptic flow
$v_2(p_T )$ for directly emitted pions (i.e. without resonance
decay contributions), comparing results for different collisions
systems and EOSs, for ideal and viscous fluid dynamics, with
parameters as indicated.
  }
 \end{figure}

Figure \ref{F1} shows the differential elliptic flow $v_2(p_T)$ for 
directly emitted pions from ideal and viscous hydrodynamics. Panels (a)
and (b) compare two systems of different size (Cu+Cu and Au+Au, both at
$b=7$\,fm), using the same equation of state (SM-EOS~Q), I-S equations
(simplified) and other free inputs. Although both systems have similar 
initial eccentricities, the smaller Cu+Cu system shows a much larger 
viscous $v_2$ suppression (by almost 70\% below the ideal fluid value 
at $p_T=2\,\mathrm{GeV}/c$ \cite{Song&HeinzPLB08,Song&HeinzPRC08-1}) than
observed in the larger Au+Au system where the suppression is almost
a factor two smaller. Panels (b) and (c) compare the same Au+Au system
at $b=7$\,fm for two different EOS and different I-S equations. Changing
the EOS from {\tt SM-EOS~Q} to {\tt EOS~L} reduces the viscous
suppression of elliptic flow by another quarter (from $\sim 40\%$ to  
$\sim 30\%$ below the ideal fluid value at $p_T=2\,\mathrm{GeV}/c$).
Replacing the simplified I-S equations used in \cite{Song&HeinzPLB08,%
Song&HeinzPRC08-1} by the full I-S equations used in \cite{Romatschke:2007mq} 
further reduces the $v_2$ suppression from 30\% to 25\% below the ideal 
hydrodynamics value at $p_T=2\,\mathrm{GeV}/c$. This final result is 
consistent with \cite{Romatschke:2007mq}. Although for {\tt EOS~L} the 
additional term in the full I-S equations only results in a 5-10\% 
difference in $v_2$ suppression, its effect is much larger for more
rapidly expanding systems, such as Cu+Cu collisions driven by a stiff
conformal EOS $e=3p$ \cite{Song&HeinzPRC08-2}. In such systems the 
simplified I-S equations lead to a strong sensitivity of physical 
observables on the value of the microscopic relaxation time $\tau_\pi$,
whereas inclusion of the last term in the full I-S equations (\ref{eq2}) 
appears to minimize this sensitivity in all situations 
\cite{Song&HeinzPRC08-2}. (It is also needed to preserve the conformal 
symmetry in conformal systems \cite{Baier:2007ix}.)

\section{Dynamical freeze-out and effects from late hadronic viscosity}
\label{sec4}
\vspace*{-2mm}

In ideal hydrodynamics, one usually imposes ``sudden freeze-out'', i.e.
a sudden transition from thermalized fluid to free-streaming particles 
on a hypersurface $\Sigma(x)$ of constant temperature or energy density
\cite{reviews}. The same algorithm has also been used in most of the 
existing viscous hydrodynamic calculations \cite{Romatschke:2007mq,%
Song&HeinzPLB08,Song&HeinzPRC08-1,Chaudhuri:2008sj,Song&HeinzPRC08-2,%
Romatschke:2008}. Since viscous hydrodynamics is based on an expansion
in small deviations from local equilibrium, its validity requires
the microscopic relaxation time to be much smaller than the macroscopic 
inverse expansion rate, $\tau_\mathrm{rel} \partial{\cdot}u \ll 1$. This
condition (whose long history is discussed in Ref.~\cite{Heinz&Greg07}
where it is also applied to ideal hydrodynamics) provides a natural 
criterium for a dynamical freeze-out algorithm. Dusling and Teaney 
\cite{Dusling&Teaney:2007gi} implemented it into their viscous 
hydrodynamics. They find that in this case the viscous $v_2$ suppression
is dominated by non-equilibrium corrections to the local thermal 
distribution function along the freeze-out surface 
\cite{Dusling&Teaney:2007gi}. This is not a collective effect arising 
from anisotropies of the flow velocity profile, but a reflection of 
non-equilibrium momentum anisotropies in the local fluid rest frame.
In contrast, for isothermal freeze-out we find \cite{Song&HeinzPRC08-1} 
that the viscous $v_2$ suppression is dominated by the viscous reduction 
of the collective flow anisotropy, while local rest frame momentum 
anisotropies play a much smaller role. This comparison shows that a 
careful treatment of the hadronic decoupling process will be required 
for the quantitative extraction of $\eta/s$ from elliptic flow data. 
Dusling also found that dynamical freeze-out can increase the slope of 
the multiplicity dependence of the eccentricity-scaled elliptic flow 
$v_2/\varepsilon$ \cite{Dusling}. This is an improvement over the 
scaling behaviour found in \cite{Song&HeinzPRC08-2} for viscous 
hydrodynamics with constant $\eta/s$ and isothermal freeze-out which
features a slope that is too small. 

There are other reasons why a proper kinetic treatment of the late 
hadronic phase is important. By matching a realistic hadron rescattering 
cascade to an ideal fluid description of the QGP and hadronization
stages, Hirano {\it et al.} \cite{Hirano:2005xf} showed that the HRG 
phase is highly viscous and strongly suppresses any buildup of elliptic
flow during the hadronic stage. This is consistent with a recent analysis
by Demir and Bass \cite{Demir:2008tr} who found large shear viscosities 
for their hadronic UrQMD cascade even close to $T_c$ (between 5 and 10 
times above the KSS bound). For a correct description of the beam
energy and centrality dependence of $v_2$, which are crucially affected
by the changing relative weight of QGP and HRG dynamics in building
elliptic flow (in central collisions or at higher energies the system 
spends more time in the QGP phase than in peripheral collisions or at
low energies), a realistic kinetic simulation of the hadronic phase
and its freeze-out thus appears to be indispensable. It will also solve
the problem of using an incorrect chemical composition in the hadronic
phase that plagues present implementations of viscous hydrodynamics: the 
empirical finding at RHIC that chemical equilibrium is broken close to 
$T_c$ and does not (as so far assumed in all viscous codes) persist to 
the point of kinetic freeze-out has important consequences for elliptic 
flow. As shown in Refs.~\cite{Hirano:2002ds,Kolb03} within the framework 
of ideal fluid dynamics, the distribution of the total momentum anisotropy 
among the various hadronic species depends strongly 
on the chemical composition of the hadronic fireball at freeze-out,
with almost 25\% larger pion elliptic flow when chemical equilibrium
is broken at $T_c\approx 165$\,MeV than for the case where pions are 
allowed to remain in chemical equilibrium down to $T_\mathrm{dec}\approx 
100$\,MeV. It is obvious that such a large effect must be correctly
implemented in viscous hydrodynamics before a quantitative extraction
of $\eta/s$ for the QGP can be attempted.    

\section{Bulk viscosity}
\label{sec5}
\vspace*{-2mm}
\begin{figure}[htb]
  \begin{center}
  \includegraphics[width=0.49\linewidth,clip=]{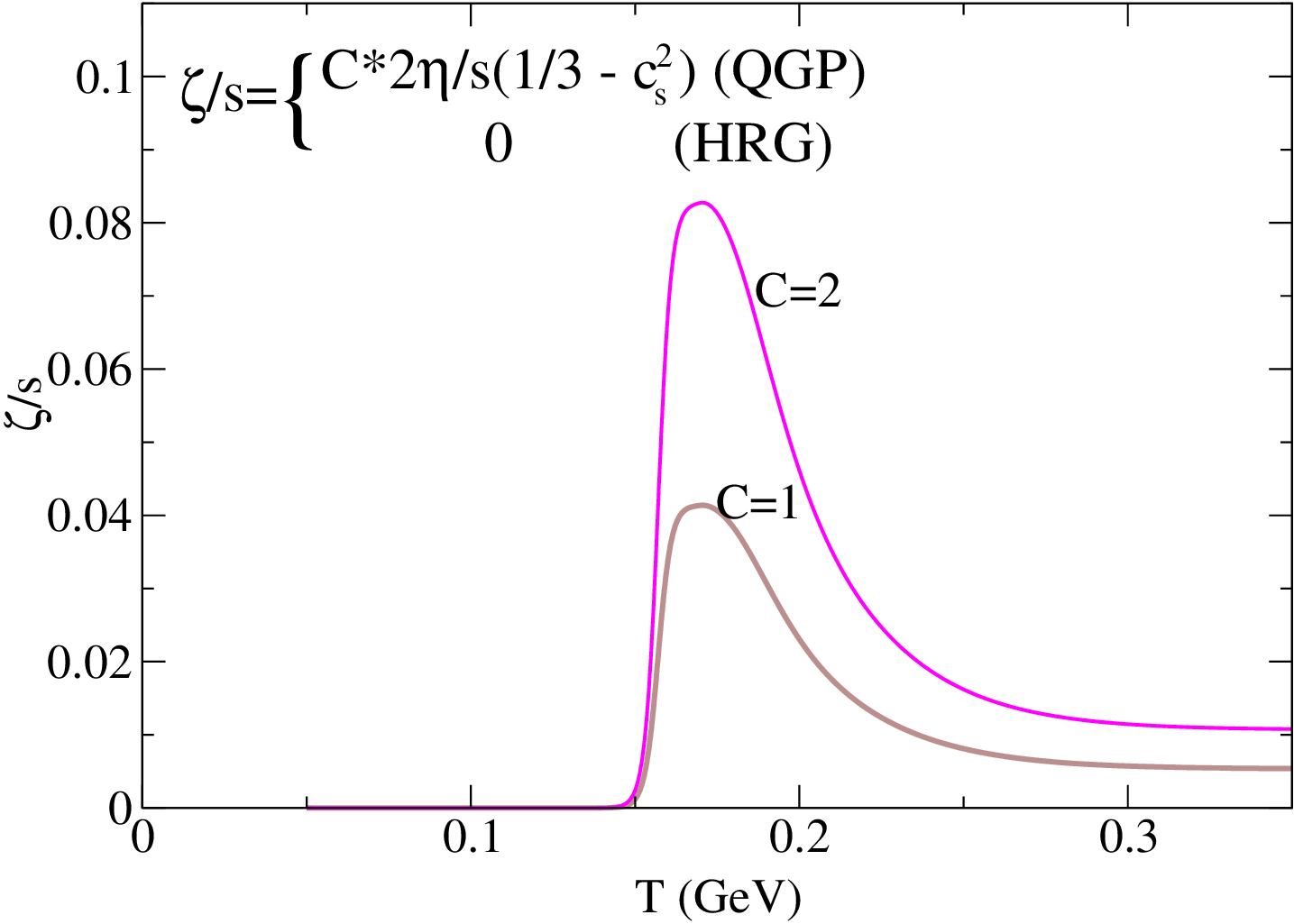}
  \includegraphics[width=0.49\linewidth,clip=]{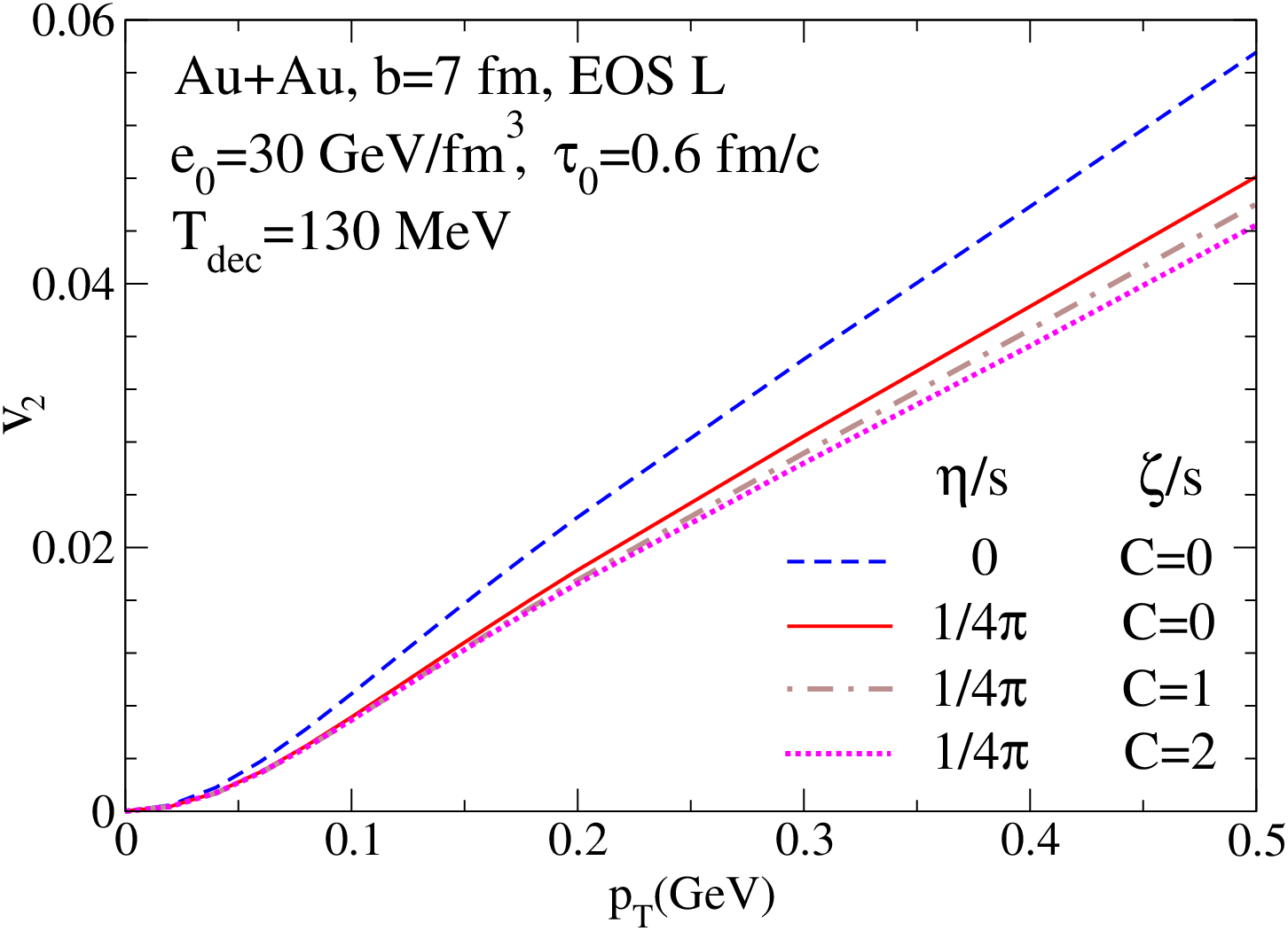}
  \end{center}
\vspace*{-2mm}
\caption{\label{F2}(Color online) {\sl Left:} The bulk viscosity to
entropy ratio as a function of temperature, as used in these proceedings
(see text for discussion). {\sl Right:} Differential elliptic
flow $v_2(p_T )$ for directly emitted pions calculated with ideal
hydrodynamics and with viscous hydrodynamics. The viscous calculations
assume minimal shear viscosity and three different normalizations
of the bulk viscosity shown in the left panel, as noted in the legend.
}
 \end{figure}
Early viscous hydrodynamic simulations ignored bulk viscosity for 
simplicity. Although it vanishes for a conformal fluid or massless QGP 
on the classical level, quantum effects break the conformal symmetry
of QCD and generate a nonzero bulk viscosity even in the massless QGP
phase, as recently measured using lattice QCD \cite{Bulk-Lattice}. 
General arguments predict that the specific bulk viscosity $\zeta/s$
exhibits a peak near $T_c$ \cite{Paech:2006st,Karsch:2007jc} and
then decreases again on the hadronic side. However, different theoretical 
approaches show dramatically different peak values: the one from AdS/CFT
predictions \cite{Bulk-AdS-1} is more than a order of magnitude smaller 
than the values extracted from lattice QCD data \cite{Bulk-Lattice}
(see, however, Ref.~\cite{Moore:2008ws} for a critical discussion of
the lattice QCD approach). Considering this theoretical uncertainty, we 
treat the bulk viscosity as an essentially free input, implementing
only the feature that it peaks around $T_c$. To obtain an explicit 
expression for $(\zeta/s)(T)$ we connect the minimum AdS/CFT result
from Ref.~\cite{Bulk-AdS-2}, evaluated with lattice QCD data for 
the speed of sound $c_s^2(T)$, through a Gaussian function peaked at 
$T_c$ with a zero value in the hadronic phase (lower brown line ($C=1$) 
in the left panel of Figure~\ref{F2}). To simulate effects from larger 
bulk viscosity, we multiply the entire function by a factor $C>1$ 
($C=2$ for the upper line in the left panel of Figure~\ref{F2}).

Since the fireball expands, the Navier-Stokes limit of the bulk 
viscous pressure $\Pi=-\zeta\theta$ is negative. This reduces the 
thermal pressure (effectively softening the EOS near $T_c$), decelerates 
longitudinal expansion and suppresses the buildup of radial flow. As a 
result, the hadron $p_T$-spectra become steeper. By construction, the 
bulk viscous effects are strongest around $T_c$, but since the transverse
edge of the fireball already hadronizes early when the longitudinal 
expansion rate is high, this does not imply that bulk viscous effects 
are necessarily small at early times. Indeed, we see in the right panel 
of Fig.~\ref{F2} that elliptic flow is significantly reduced by bulk 
viscosity even though a large fraction of the momentum anisotropy is 
generated well before most of the matter reaches $T_c$. For minimal
shear viscosity $\eta/s=1/4\pi$, we find that bulk viscosity can increase
the viscous suppression of pion elliptic flow at $p_T=0.5$\,GeV by an 
additional 25\% (for $C=1$) to 50\% (for $C=2$) relative to the case 
$\zeta=0$.

Clearly, bulk viscous effects must be accounted for when trying to
extract $\eta/s$ from elliptic flow data. At this moment it is not clear
which observables can be used for a clean separation of bulk and shear
viscous effects. One should explore whether the centrality dependence 
of the local hydrodynamic expansion rate impacts radial and elliptic 
flow in distinct ways that allow for such a separation.   

\section{Glauber model vs. CGC initialization}
\label{sec6}
\vspace*{-2mm}

We now come to what may turn out to be a serious road block for precision
measurements of the QGP shear viscosity: our insufficient knowledge of
the initial source eccentricity $\varepsilon$. It has now been known for 
while that the Color Glass Condensate (CGC) model, implemented in the 
initial entropy or energy density profile via the KLN parametrization (see 
\cite{Hirano:2005xf,Drescher:2006pi} for references), leads to $\sim 30\%$ 
larger initial source eccentricities than the popular Glauber model. Ideal 
fluid dynamics transforms this larger source eccentricity into $\sim30\%$
larger elliptic flow. Since the extraction of $\eta/s$ is based on 
the viscous {\em suppression} of $v_2$, obtained by comparing the 
measured elliptic flow with an ideal (inviscid) fluid dynamical benchmark
calculation, a 30\% uncertainty in this benchmark can translate into a
100\% uncertainty in the extracted value for $\eta/s$. This was recently
shown by Luzum and Romatschke (see Fig.~8 in \cite{Romatschke:2008}).
This uncertainty trumps most of the other uncertainties discussed above.
Worse, since the initial source eccentricity depends on details of the 
shape of the KLN profile near the edge of the distribution where the
gluon saturation momentum scale $Q_s$ becomes small and the CGC model 
reaches its limit of applicability, there is little hope that we can 
eliminate it theoretically from first principles.

Based on their analysis of charged hadron elliptic flow data from the 
STAR experiment, allowing for a 20\% systematic uncertainty of these 
data, the authors of \cite{Romatschke:2008} found an allowed
range $0 < \eta/s < 0.1$ for Glauber and $0.08 < \eta/s < 0.2$ for CGC 
initial conditions. Since their analysis did not include a comprehensive 
investigation of effects caused by permissible variations of the EOS 
near $T_c$, by bulk viscosity, or by late hadronic viscosity and 
non-equilibrium chemical composition at freeze-out (see Secs. 
\ref{sec3} -- \ref{sec5} above), one should add a significant additional 
uncertainty band to these ranges. Furthermore, correcting the experimental 
data for event-by-event fluctuations in the initial source eccentricity 
\cite{Poskanzer} may bring down the measured $v_2$ values even below the 
range considered in \cite{Romatschke:2008}. Still, we agree with Luzum 
and Romatschke that, even when adding all the above effects in magnitude
(ignoring the fact that several of them clearly have opposite signs), 
viscous hydrodynamics with $\eta/s > 5 \times (1/4\pi)$ would suppress 
the elliptic flow too much to be incompatible with experiment. 

\section{Numerical viscosity}
\label{sec7}
\vspace*{-2mm}

To study the effects from shear and bulk viscosity one must ensure 
that numerical viscosity is under control and sufficiently small. 
Simply speaking, numerical viscosity comes from the discretization 
of the hydrodynamic equations for numerical calculation. It causes 
entropy production even in ideal hydrodynamics without shocks and can 
never be fully avoided. To minimize numerical viscosity, the flux-corrected 
transport algorithm {\tt SHASTA} \cite{SHASTA} employed by {\tt VISH2+1} 
(and by its ideal fluid ancestor {\tt AZHYDRO} \cite{AZHYDRO})
implements an ``antidiffusion step'' involving a parameter $\Lambda$ called 
``antidiffusion constant'' \cite{SHASTA}. For a given grid spacing, numerical 
viscosity is maximized by setting $\Lambda=0$. In standard situations, the 
default value $\Lambda=\frac{1}{8}$ minimizes numerical viscosity effects 
\cite{SHASTA}. With $\Lambda=\frac{1}{8}$ and typical grid spacing $\Delta
x=\Delta y=0.1$\,fm, $\Delta \tau=0.04$\,fm/$c$, {\tt AZHYDRO} generates 
only 0.3\% additional entropy in central Au+Au collisions. This is
negligible when compared with the ${\cal O}(10\%)$ entropy production 
by {\tt VISH2+1} for a fluid with real shear viscosity $\eta/s=1/4\pi$.

By increasing the grid spacing in {\tt AZHYDRO} and/or changing $\Lambda$,
we can explore the effects of numerical viscosity on radial and elliptic 
flow. We find that numerical viscosity has little effect on the development
of radial flow but reduces $v_2$ in very much the same way as does real 
shear viscosity. Since we gauge the effects of $\eta/s$ on $v_2$ by comparing
results from {\tt VISH2+1} for $\eta/s\ne 0$ to those for $\eta/s=0$,
we should explore how much in the latter case $v_2$ is already suppressed
by numerical viscosity. We can do this by setting $\eta/s=0$ and reducing 
the grid spacing until $v_2$ stops changing (i.e. until we have completely 
removed all numerical viscosity effects on $v_2$). In this way we have 
ascertained that for our standard grid spacing numerical viscosity 
suppresses the differential elliptic flow $v_2(p_T)$ by less than 2\%. 

\section{Summary and outlook}
\label{sec8}
\vspace*{-2mm}

While the elliptic flow $v_2$ generated in non-central heavy-ion collisions 
is very sensitive to the shear viscosity to entropy ratio $\eta/s$ of the
QGP, it is also significantly affected by (i) details of the initialization 
of the hydrodynamic evolution, (ii) bulk viscosity and sound speed near the 
quark-hadron phase transition, and (iii) the chemical composition and 
non-equilibrium kinetics during the late hadronic stage. Not all of these 
effects are presently fully under control. Recent attempts to extract the 
specific shear viscosity $\eta/s$ phenomenologically, by comparing 
experimental elliptic flow data with viscous hydrodynamics, have 
established a robust upper limit
\begin{eqnarray}
\label{etas}
  \left.\frac{\eta}{s}\right|_\mathrm{QGP} < 5\times \frac{1}{4\pi},
\end{eqnarray}
close to the conjectured KSS bound \cite{Kovtun:2004de}, but further 
progress requires elimination of the above systematic uncertainties. 
Since some of these influence the build-up of elliptic flow in opposite 
directions, it is quite conceivable that the QGP specific viscosity is 
in fact much closer to the KSS bound $\left.(\eta/s)\right|_\mathrm{KSS} 
=1/4\pi$ than suggested by the upper limit (\ref{etas}). Ongoing 
improvements on the theory side should help to reduce or eliminate most 
of the above uncertainties, bringing us closer to a quantitative 
extraction of $\eta/s$ for the quark-gluon plasma. The single largest 
uncertainty, however, is caused by our poor knowledge of the initial 
source eccentricity which varies by about 30\% between models. As shown 
in \cite{Romatschke:2008}, this translates into an ${\cal O}(100\%)$ 
uncertainty for $\eta/s$. It seems unlikely that theory can help to 
eliminate this uncertainty from first principles. It thus appears crucial 
to develop experimental techniques that may help us to pin down the 
initial source eccentricity phenomenologically, with quantitative 
precision at the percent level.  

\section*{Acknowledgments:} 
\vspace*{-2mm}
We thank K. Dusling, R. Fries, P. Huovinen, M. Lisa, P. Petreczky, and 
X.-N. Wang for fruitful discussions. This work was supported by the U.S. 
Department of Energy under grant DE-FG02-01ER41190.


\end{document}